\documentclass[
pre,
preprint,
tightenlines,
superscriptaddress,
amsmath,
amssymb,
noshowkeys,
nofootinbib,
floatfix,
]{revtex4}
 \usepackage{color}
\usepackage{framed}
\usepackage{subfigure}
\usepackage{time}
\usepackage{graphicx}
\usepackage{enumerate}
\usepackage{latexsym}
\usepackage{bm}
\usepackage{upgreek}
\usepackage{amsthm}
\usepackage{amssymb}
\usepackage{mathrsfs}
\usepackage{amsmath}
\usepackage{physics} 

\usepackage{graphicx}          
\usepackage{subfigure}         
\usepackage{dcolumn}           
\usepackage{capt-of}           
\usepackage{makecell}          
\usepackage{bbm}               
\usepackage{bm}                
\usepackage{time}              
\usepackage[mathlines]{lineno} 
\usepackage{stackrel}          
\usepackage{enumitem}          
\usepackage{physics}           
\usepackage[mathscr]{eucal}    
\usepackage{hyperref}          
\hypersetup{
    pdfnewwindow=true,         
    colorlinks=true,           
    linkcolor=red,             
    citecolor=blue,            
    filecolor=green,           
    urlcolor=cyan              
}
\usepackage{amsthm}
\theoremstyle{plain}                           

\theoremstyle{definition}

\theoremstyle{remark}

\newcommand{\rmi}{{\rm i}}                     
\newcommand{\rme}{{\rm e}}                     
\newcommand{\Partial}[4]
   {\Bigl ( \frac{\partial #1 }{\partial #2 } \Bigr )_{\! #3, #4 }}
\newcommand{\Tproduct}[1]%
   {\ensuremath{\mathcal{T} \{ \, #1 \, \} } }
\DeclareSymbolFont{matha}{OML}{txmi}{m}{it} 
\DeclareMathSymbol{\varv}{\mathord}{matha}{118}
\newcommand{\be}{\begin{equation}}
\newcommand{\ee}{\end{equation}}
\newcommand{\bea}{\begin{eqnarray}}
\newcommand{\eea}{\end{eqnarray}}

\begin{document}
%
%
\preprint{LA-UR-25-25606}
\title{Solitary waves in the complementary generalized ABS model}
\author{Avinash Khare}
\email[Email: ]{avinashkhare45@gmail.com} 
\affiliation{
   Physics Department, 
   Savitribai Phule Pune University, 
   Pune 411007, India} 
\author{Fred Cooper}
\email[Email: ]{cooper@santafe.edu}
\affiliation{
   Santa Fe Institute,
   1399 Hyde Park Road,
   Santa Fe, NM 87501, USA}
\affiliation{
   Center for Nonlinear Studies and Theoretical Division, 
   Los Alamos National Laboratory, 
   Los Alamos, NM 87545, USA}
\author{John F. Dawson}
\email[Email: ]{john.dawson@unh.edu}
\affiliation{
   Department of Physics,
   University of New Hampshire,
   Durham, NH 03824, USA}
\author{Avadh Saxena} 
\email[Email: ]{avadh@lanl.gov} 
\affiliation{
   Center for Nonlinear Studies and Theoretical Division, 
   Los Alamos National Laboratory, 
   Los Alamos, NM 87545, USA}
\date{\today, \now \ PST}
\begin{abstract}
We obtain exact solutions of the nonlinear Dirac equation in 1+1 dimension of  
the form  $  \Psi(x,t) =\Phi(x) \rme^{-\rmi \omega t}$ where the nonlinear 
interactions are a combination of vector-vector and scalar-scalar  
interactions with the interaction Lagrangian given by 
$L_I = \frac{g^2}{(\kappa+1)}[\bar{\psi} \gamma_{\mu}\psi 
\bar{\psi} \gamma^{\mu} \psi]^{(\kappa+1)/2} - \frac{g^2}{q(\kappa+1)}(\bar{\psi} \psi)^{\kappa+1}$, 
where $\kappa>0$ and $q>1$. This is the complement of the generalization of the  ABS  
model \cite{abs}   that we recently studied \cite{ak} and denoted as the 
gABS model.  We show that like the gABS model, in the complementary gABS 
models the solitary wave solutions also exist in the entire $(\kappa, q)$ plane and 
further in both models  energy of the solitary wave divided by its charge is
 {\it independent} of the coupling constant $g$. However, unlike the gABS model 
here all the solitary waves are single humped, any value of $0 < \omega < m$ is
allowed and further unlike the gABS 
model, for this complementary gABS model the solitary wave bound states exist 
only in case $\kappa \le \kappa_c$, where $\kappa_c$ depends on the value of 
$q$. Here $\omega$ and $m$ denote frequency and mass, respectively. We discuss 
the regions of stability of these solutions as a function of $\omega,q,\kappa$ 
using the Vakhitov-Kolokolov criterion. 
Finally we discuss the non-relativistic reduction of the two-parameter family
of  this complementary generalized ABS model to a modified nonlinear 
Schr\"odinger equation (NLSE) and discuss the stability of the solitary waves 
in the domain of validity of the modified NLSE.
\end{abstract}
\maketitle
%
\section{\label{s:Intro}Introduction}

The Dirac equation is one of the most important elements of the modern gauge
theories. After the seminal paper of Dirac and others, the nonlinear Dirac 
equation with  scalar-scalar (SS) as well as vector-vector (VV) 
interactions was introduced by Ivanenko \cite{Iv1} and Thirring \cite{R3},  
respectively. In the 1970s, these models were extensively applied in the context of 
particle physics; for example to describe extended nucleons \cite{R4} or to explain
quark confinement \cite{R5}. Subsequently, these models have found applications
in the context of optical gratings \cite{gratings}. Other applications of the nonlinear 
Dirac equation are related to the light propagation in honeycomb photo-refractive 
lattices such as photonic graphene \cite{R10}, conical diffraction in such 
structures \cite{R11}, spin-orbit coupled Bose-Einstein condensates \cite{R12} 
and in the context of polarons in conducting polymers \cite{bc}. We note that Serge 
Aubry, whom we honor in this paper and special issue, contributed significantly to 
the understanding of nonlinear equations and nonlinear lattices \cite{aubry}. 

The explicit solutions of the $1+1$ dimensional Soler model as well as Thirring
model were known since 1974 \cite{lee}. In 2010 these one dimensional models
were generalized to arbitrary nonlinearity characterized by a parameter $\kappa$ 
and explicit solitary wave solutions were obtained for arbitrary $\kappa > 0$ \cite{fred}. 
Further, it was shown that whereas the solitary wave bound states exist for any 
$\kappa > 0$, in the scalar-scalar interaction case, in the vector-vector
case it was shown that the solitary wave bound states only exist if 
$\kappa < \kappa_c$. Recently, Alexeva, Barashenkov and Saxena \cite{abs} 
introduced a novel model at $\kappa = 1$ with an admixture of both the SS and  
the VV interactions. Recently, we \cite{ak} have generalized this model to a two
(continuous) parameter family of models (which we call hereafter as the gABS 
model) characterized by the nonlinearity parameter $\kappa > 0$ as well as a 
parameter $p > 1$ characterizing the strength of the VV interaction vis a vis the 
SS interaction and showed that the solitary wave bound states exist in the entire 
$\kappa$-$p$ plane. The interaction Lagrangian for the gABS model is given by \cite{ak}
\be\label{a}
L_I =({g^2}/{(\kappa+1)})(\bar{\psi} \psi)^{\kappa+1}-({g^2}/{p(\kappa+1)})
[\bar{\psi} \gamma_{\mu} \psi \bar{\psi} \gamma^{\mu} \psi]^{(\kappa+1)/2}\,, 
\ee
where the nonlinearity parameter $\kappa > 0$ while VV strength parameter is
characterized by $p > 1$. 
It is worth emphasizing that in the gABS model the VV interaction term has
negative sign compared to the SS term which has positive sign. It is then 
natural to enquire if one can also construct a {\it complementary} gABS model in 
which the SS interaction term has negative sign while the VV interaction 
term has positive sign. The purpose of this paper is to introduce such a 
complementary gABS model and compare and contrast its various features with 
those of the gABS model. The interaction Lagrangian for the complementary 
gABS model is given by
\be\label{b}
L_I = \frac{g^2}{(\kappa+1)}[\bar{\psi} \gamma_{\mu}\psi 
\bar{\psi} \gamma^{\mu} \psi]^{(\kappa+1)/2} 
- \frac{g^2}{q(\kappa+1)}(\bar{\psi} \psi)^{\kappa+1}\,,
\ee
where the nonlinearity parameter $\kappa > 0$ while the SS strength parameter 
is characterized by $q > 1$. 

The plan of the paper is the following. In Sec. II we obtain the rest frame
solitary wave solutions of the form $\psi(x,t) = e^{-i\omega t} \psi(x)$ 
which are functions of the rest frame frequency $\omega$ as well as $q, \kappa$ and
$g$. In this paper we are interested in looking for solutions for which 
$0 < \omega < m$. In Sec. III we discuss the various properties of the solitary
wave solutions. We show that unlike the gABS model but like the pure VV or SS 
cases, in the complementary gABS model, the rest frame frequency can take any 
value in the range $0 < \omega < m$. Further, we show that unlike the SS or the
gABS models but like the VV case, in the complementary gABS model the charge
density always has single hump behavior in the entire allowed range of 
$\omega$. Further, we calculate the total energy $E$ as well as the charge $Q$
and show that while $E$ and $Q$ depend on the coupling constant $g$, their 
ratio $E/Q$ is independent of $g$. However, unlike the SS and the gABS models
but like the VV case we find that in the complementary gABS model, solitary 
wave bound states are not allowed for all the values of $\kappa$ ($> 0$). In
particular, in the complementary gABS model the bound states are allowed only
if $\kappa \le \kappa_c$ with $\kappa_c$ depending on $q$ 
(note $1 < q < \infty$). In Sec. IV we study the stability of the allowed 
solitary wave bound states using the Vakhitov-Kolokolov \cite{vk} criterion and
show the stability of these solitary wave bound states in case $\kappa \le 2$. 
In Sec. V we discuss the nonrelativistic reduction of the complementary gABS 
model to the generalized modified nonlinear Schr\"odinger equation (NLSE) and 
discuss the stability of the solitary waves assuming the validity of the modified NLSE.
In Sec. VI we summarize the main results obtained in this paper and point out some
of the open problems.

\section{ Nonlinear Dirac Equation With VV-SS Interaction}

The generalization of the ABS model \cite{abs}, which we dubbed  gABS \cite{ak}, has as an interaction Lagrangian:
$L_I =({g^2}/{(\kappa+1)})(\bar{\psi} \psi)^{\kappa+1}-({g^2}/{p(\kappa+1)})
[\bar{\psi} \gamma_{\mu} \psi \bar{\psi} \gamma^{\mu} \psi]^{(\kappa+1)/2}$.   When $p=2, \kappa=1$, this model reduced to the original ABS model, which had an additional supersymmetry. We determined the solitary wave solutions of this equation and studied their properties as a function of $\kappa, p,\omega$, where $\omega$ is the frequency of the static solution. 
Here we are considering the corresponding complementary model with VV-SS interaction, with a parameter $q> 1$, which reduces to the VV  model for large $q$. 
 
The Lagrangian density for the generalized VV-SS model is of the form:
\be\label{1}
L = i\bar{\psi} (\gamma^{\mu} \partial_{\mu} -m) \psi + L_I\,,
\ee
where
\be\label{2}
L_I = \frac{g^2}{(\kappa+1)}[\bar{\psi} \gamma_{\mu}\psi 
\bar{\psi} \gamma^{\mu} \psi]^{(\kappa+1)/2} 
- \frac{g^2}{q(\kappa+1)}(\bar{\psi} \psi)^{\kappa+1} \,. 
\ee 
\subsection{ Solitary Wave Solutions}

We will follow the notations of our 2010 PRE paper \cite{fred}. It turns out 
that the several basic results given there are also valid for our case.
We are interested in finding the solitary wave solutions of the generalized  
model characterized by the Lagrangian as given by Eqs. (\ref{1}) and (\ref{2}).
The corresponding field equation is
\be\label{3}
(\gamma^{\mu} \partial_{\mu} - m) \psi+g^2 [(\bar{\psi}\gamma_{\mu} \psi)
(\bar{\psi} \gamma^{\mu} \psi)]^{(\kappa-1)/2}\psi 
- \frac{g^2}{q} (\bar{\psi} \psi)^{\kappa} \psi \,. 
\ee 
We  will use the same  representation of gamma matrices as used in  \cite{fred}:
\be\label{4}
\gamma^{0} = \sigma_3\,,~~ \gamma^{1} = i \sigma_1\,,
\ee
and $\gamma_5 = \sigma_2$, where $\sigma_i$ are the Pauli matrices. The gamma 
matrices then obey the 
anti-commutation relation, $[\gamma^{\mu},\gamma^{\nu}]_{+} = 2 g^{\mu\nu}$.

For our generalized Eq. (\ref{3}) we look for solutions (in the solitary 
wave rest frame) of the form,
\be\label{5}
\psi(x,t) = (u(x), v(x)) \, e^{-i \omega t}
= R(x) (\cos(\theta(x)), \sin(\theta(x))) \, e^{-i \omega t} \,. 
\ee
It is easy to check the current conservation
\be\label{6}
\partial_{\mu} j^{\mu}(x) = 0\,,~~j^{\mu} = \bar{\psi} \gamma^{\mu} \psi\,.
\ee
This implies that the charge $Q$ is independent of time, where
\be\label{7}
 Q =\int dx\, j^{0}(x) =\int dx\, \psi^{\dag} \psi\,.
\ee
Further, the stress-energy tensor is also conserved:
\be\label{8}
 \partial_{\mu} T^{\mu\nu}(x) = 0\,,~~
 T_{\mu\nu} = \frac{1}{2}[D_{\mu\nu} + {h.c.}] - g_{\mu\nu} L\,,~~ 
D_{\mu\nu} = \bar{\psi} \gamma_{\mu} \partial_{\nu} \psi \>,   
\ee
which means that the linear momentum vector $P^{\nu} = (E,{P^{i}})$ is 
conserved:
\be\label{9}
P_{\nu} = \int dx\,  T_{0 \nu}(x)\,.
\ee
For stationary solutions  this leads to 
\be\label{10}
T_{10}= constant;~~ T_{11}= constant\,.
\ee
We can write
\be\label{11}
T_{11}= \omega \psi^\dag \psi - m \bar{\psi} \psi + L_I\,. 
\ee
For solitary wave solutions vanishing at infinity the constant is zero so that
\be\label{12} 
\omega \psi^\dag \psi - m \bar{\psi} \psi + L_I =0\,.
\ee
On multiplying Eq. (\ref{3}) on the left by $\bar{\psi}$ we have 
\be\label{13}
(\kappa+ 1) L_I = -\omega \psi^{\dag} \psi + m \bar{\psi} \psi 
+ i \bar{\psi} \gamma_1 \partial_1 \psi\,.
\ee
Thus
\be\label{14}
\omega \kappa \psi^{\dag} \psi - m \kappa  \bar{\psi} \psi 
+ i\bar{\psi} \gamma_1 \partial_1 \psi =0\,.
\ee
We also have 
\be\label{15}
\kappa L_I =   i \bar{\psi} \gamma_1 \partial_1 \psi \,, 
\ee
\be\label{16}
\bar{\psi} i\gamma_1 \partial_1 \psi = \psi^{\dag} \psi \frac{d\theta}{dx}\,.
\ee
This leads to the simple differential equation for $\theta$
\be\label{17}
\frac{d\theta}{dx} = \kappa [m\cos(2\theta) -\omega]\,,
\ee
whose solution is 
\be \label{18}
\tan(\theta) = \alpha \tanh(\kappa \beta x) \,, 
\ee
with 
\be\label{19}
\beta = \sqrt{m^2 - \omega^2}\,,~~\alpha = \sqrt{\frac{m-\omega}{m+\omega}}\,.
\ee

\begin{figure} [h]  
 \includegraphics[width=0.7 \linewidth]{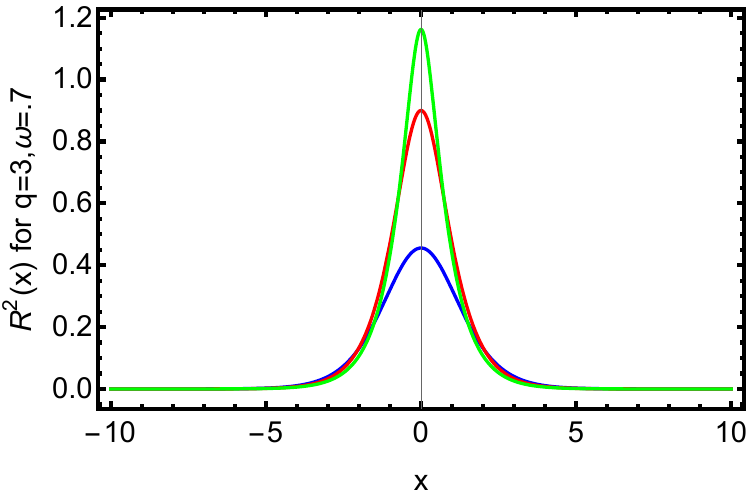}
    \caption{ Plot of $R^2$ for $q=3,\omega=0.7$ at $\kappa=0.5$ (blue), $\kappa=1$ (red) and $\kappa=2$ (green).}
     \label{fig1} 
    \centering
\end{figure}

In terms of $R, \theta$ we have that
\be\label{20}
L_I = \frac{g^2}{\kappa+1}R^{2\kappa}[1-\frac{1}{q}\cos^{\kappa+1}(2\theta)]\,,
\ee
so that using Eq. (\ref{11}) we obtain
\be\label{21}
R^2[\omega-m\cos(2\theta)]+\frac{g^2 R^{2(\kappa+1)}}{\kappa+1}
[1 - \frac{1}{q}\cos^{\kappa+1}(2\theta)] = 0\,.
\ee
This leads to
\be\label{22} 
R^2= \left[\frac{(\kappa+1) (m \cos 2 \theta - \omega)}
{g^2 \big (1 - \frac{1}{q} \cos(2\theta)^{\kappa+1} \big )}\right]^{1/\kappa}\,.
\ee
As expected, in the limit $q \rightarrow \infty$, $R^2$ reduces to that of the
VV case \cite{fred}. On using Eqs. (\ref{18}) and (\ref{19}) and the identities
\be\label{23}
m+\omega\cosh(2\kappa \beta x) = \frac{(m+\omega)
[1-\alpha^2 \tanh^2(\kappa \beta x)]}{\sech^2(\kappa \beta x)}\,,
\ee
\be\label{24}
\omega+m\cosh(2\kappa \beta x) = \frac{(m+\omega)
[1+\alpha^2 \tanh^2(\kappa \beta x)]}{\sech^2(\kappa \beta x)}\,,
\ee
one obtains an alternative expression for $R^2$ 
\be\label{25}
R^2= (1+\alpha^2 y^2) \bigg[\frac{(\kappa+1)(1-y^2) (m-\omega)}{g^2 
[(1+ \alpha^2 y^2)^{\kappa+1}
-\frac{1}{q}(1-\alpha^2 y^2)^{\kappa+1}]} \bigg]^{1/\kappa} \,,  
\ee
where $y= \tanh (\kappa \beta x)$.
In Fig.~(\ref{fig1}), we show $R^2$ at  $q=3,\omega=0.7$ at $\kappa=0.5$ (blue), $\kappa=1$ (red) and $\kappa=2$ (green). 

\section{Properties of the Solitary Wave Solutions}

In this section we discuss the various properties of the solitary wave 
solutions obtained above.

\subsection{Evaluation of Charge $Q$ and Range of $\omega$}

It is straightforward to compute the charge $Q$ using Eqs. (\ref{7}) and 
(\ref{25}). On making the substitution $y = \tanh(\beta x)$ we obtain
\be\label{3.1}
Q = C(\omega,\kappa) I(\omega,\kappa,q)\,,
\ee
where
\be\label{3.1a}
C(\omega,\kappa) = \frac{2}{\kappa \beta}[\frac{(\kappa+1)(m-\omega)}
{g^2}]^{1/\kappa}\,, 
\ee
while
\be\label{3.2}
I(\omega,\kappa,q) = \int_{0}^{1} dy~ \frac{(1+ \alpha^2 y^2)
(1-y^2)^{(1-\kappa)/\kappa}}{[+(1+\alpha^2 y^2)^{\kappa+1}
-\frac{1}{q}(1-\alpha^2 y^2)^{\kappa+1}]^{1/\kappa}}\,.
\ee
One finds that the integral always converges at $y = 1$
so that entire range of $0 < \omega < m$ is allowed.  This is to be contrasted
with the gABS model \cite{ak} where $\omega$ is restricted to 
\be\label{1.4}
\frac{1}{p^{1/(\kappa+1)}} < \frac{\omega}{m} < 1\,.
\ee

\subsection{Single Hump Behavior of $R^2$}

\begin{figure}[h]
    \centering
    \includegraphics[width=0.7 \linewidth]{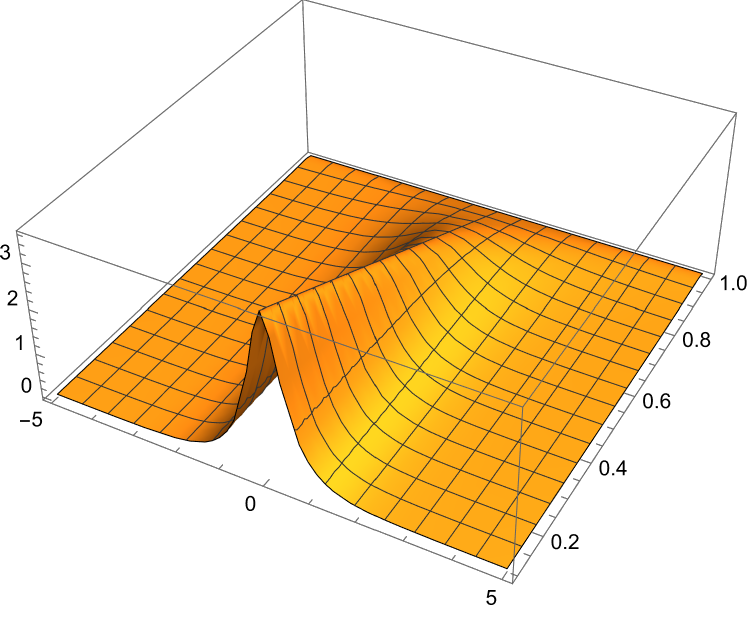}
    \caption{Plot of $R^2 (x,\omega)$ when $\kappa=1$ and $q=2 $.}
     \label{1hump} 
\end{figure}

For the gABS model \cite{ak} we found that the shape of the solitary wave
changes from the double humped to the single humped as one increases $\omega$. 
We now show that in contrast to the gABS model, in the complementary gABS 
model the solitary waves are always single humped.
To this end we rewrite Eq. (\ref{25}) as
\be\label{3.4}
R^2/C = (1+\alpha^2 T^2)\frac{S^{2/\kappa}}
{[(1+\alpha^2 T^2)^{\kappa+1}]^{1/\kappa}
-\frac{1}{q}(1-\alpha^2 T^2)^{\kappa+1}}\,, 
\ee
where $T$ and $S$ stand for $\tanh(\beta x)$ and $\sech(\beta x)$, respectively  
while in $C \equiv C(\omega, \kappa)$ defined in Eq. \eqref{3.1a} we have absorbed all inessential $x$ independent constants.
Hence we have
\bea\label{3.5}
&&\frac{dR^2(x)/C\kappa \beta}{dx} = 2T S^{2/\kappa} \frac{[\alpha^2 S^{2} 
-(1/\kappa)(1+\alpha^2 T^2)]}{[(1+\alpha^2 T^2)^{\kappa+1}-\frac{1}{q} 
(1-\alpha^2 T^2)^{(\kappa+1)}]^{1/\kappa}}
\nonumber \\
&&-\frac{2(\kappa+1)}{\kappa} S^{2(\kappa+1)/\kappa}(1+\alpha^2 T^2) 
\nonumber \\
&&\times\frac{\frac{1}{q}\alpha^2 T(1-\alpha^2 T^2)^{\kappa}+T\alpha^2
(1+\alpha^2 T^2)^{\kappa}}{[-\frac{1}{q}(1-\alpha^2 T^2)^{\kappa+1}
+(1+\alpha^2 T^2)^{\kappa+1}]^{(\kappa+1)/\kappa}}\,.
\eea
One of the minima of $R^2$ is thus at $\sech(\beta x) = 0$, i.e. at 
$x = \pm \infty$. On the other hand, it turns out that $x = 0$ is always a 
maximum of $R^2$ so that $R^2$ always has a single hump.
A typical behavior  of the soliton as we keep $q,\kappa$ fixed and vary omega is shown in Fig. (\ref{1hump}).

\subsection{The Total Energy $E$ and $g$ Independence of $E/Q$}
	
The Hamiltonian density in the complementary generalized ABS model is given by
\be\label{3.12}
h = T_{00} =  i \bar{\psi} \gamma_1 \partial_1 \psi+ m \bar{\psi} \psi -L_I
\equiv  h_1+h_2-h_3\,.
\ee
But from Eq. (\ref{16}) it follows that $h_3=h_1/\kappa$, and hence
\be\label{3.13}
h= h_1\left(1-\frac{1}{\kappa}\right)+h_2.
\ee
Thus the total energy $E$ is given by
\be\label{3.14}
E = \left(1-\frac{1}{\kappa}\right) H_1 + H_2\,,
\ee
where in view of Eqs. (\ref{17}) and (\ref{18}) one can write
\be\label{3.15}
H_1 = \int_{-\infty}^{\infty} dx\, R^2(x) \frac{d\theta}{dx}  
= \kappa \int_{-\infty}^{\infty} dx\, R^2(x) [m\cos(2\theta)-\omega]\,, 
\ee
while $H_2$ is given by
\be\label{3.16}
H_2 = \int_{-\infty}^{\infty} dx\, R^2(x) \cos(2\theta)\,.
\ee 
On using Eqs. (\ref{9}), (\ref{3.15}) and (\ref{3.16}) one obtains a novel 
relation between $H_1$ and $H_2$ given by
\be\label{3.17}
H_1 = \kappa H_2 - \kappa \omega Q\,,
\ee
and hence we can reexpress the ratio $E/Q$ in terms of $H_2/Q$, i.e. we obtain
\be\label{3.18}
\frac{E}{Q} = \frac{\kappa H_2}{Q} + (1-\kappa) \omega\,.
\ee

On using Eqs. (\ref{18}), (\ref{19}) and (\ref{25}) we obtain
\be\label{3.19}
H_2 =  \frac{2m}{\kappa \beta}\left[\frac{(\kappa+1)(m-\omega)}{g^2}\right]^{1/\kappa}  
J(\omega,\kappa,p) \,, 
\ee
where
\be\label{3.20}
J(\omega,\kappa,q) = \int_{0}^{1} dy~ \frac{(1- \alpha^2 y^2)
(1-y^2)^{(1-\kappa)/\kappa}}{[-\frac{1}{q}(1- \alpha^2 y^2)^{\kappa+1}
+(1+\alpha^2 y^2)^{\kappa+1}]^{1/\kappa}}\,.
\ee
Notice that $H_2 \propto (g^2)^{-1/\kappa}$ and since as shown above
$Q$ is also $\propto (g^2)^{-1/\kappa}$, thus remarkably $H_2/Q$ and hence
$E/Q$ is independent of $g^2$ and only depends on $\omega, \kappa$ and $p$.
Unfortunately, $Q$ and $H_2$ have to be computed numerically except when
$\kappa = 1$.

\subsection{$E$ and $Q$ at $\kappa = 1$} 

When $\kappa = 1$, $R^2$ as given by Eq. (\ref{23}) takes the simpler form 
\be\label{3.21} 
R^2= \frac{2 (m \cos 2 \theta - \omega)}{g^2 [1-\frac{1}{q} \cos^2(2\theta)]}\,,
\ee
which can be re-expressed as
\be\label{3.22} 
R^2= \frac{2 (m \cos 2 \theta - \omega)}
{g^2 [1-\frac{1}{q}\cos^2(2\theta)]}\,.
\ee

On using Eqs. (\ref{3.1}) to (\ref{3.2}), at $\kappa = 1$, $Q$ is given by
\be\label{3.24}
Q = \frac{4\alpha}{g^2} \int_{0}^{1} dy\, \frac{(1+\alpha^2 y^2)}
{-\frac{1}{q}(1- \alpha^2 y^2)^{2} +(1+\alpha^2 y^2)2}\,.
\ee
On using partial fractions, one finds that 
\be\label{3.25}
Q = \frac{2\sqrt{q}}{g^2\sqrt{(q-1)}} 
\tan^{-1}\left(\frac{\beta\sqrt{q}}{\omega\sqrt{(q-1)}}\right)\,.
\ee
Similarly, on using Eqs. (\ref{3.19}) and (\ref{3.20}), $E = H_2$ is given by
\be\label{3.26}
E = H_2 = \frac{4m\alpha}{g^2} \int_{0}^{1} dy\, \frac{(1-\alpha^2 y^2)}
{-\frac{1}{q}(1- \alpha^2 y^2)^{2}+(1+\alpha^2 y^2)2}\,.
\ee
On using partial fractions, it is easily evaluated and we find that 
\be\label{3.27}
E = H_2 = \frac{2mq}{g^2 \sqrt{(q-1)}} 
\tan^{-1} \left (\frac{\beta}{m\sqrt{(q-1)}} \right )\,.
\ee

\begin{figure}[t]
    \centering
    \includegraphics[width=0.7 \linewidth]{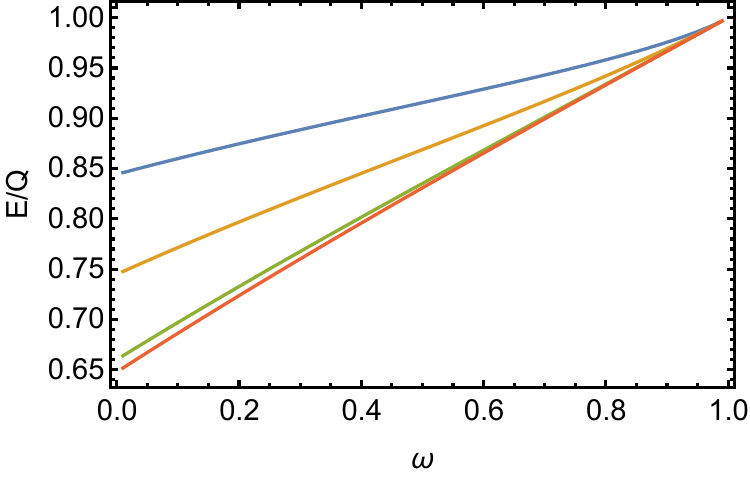}
    \caption{\label{EQk1}  Plot of $E(\omega,q)/Q(\omega,q)$ when  $\kappa=1$ for $q=1.1$ (red), 1.5 (green), 5 (yellow),10 (blue). }
\end{figure}

Thus for $m = 1$ so that $0 < \omega < 1$, we have 
\be\label{3.28}
E/Q = \sqrt{q} \frac{\tan^{-1}(\beta/\sqrt{q-1})}{\tan^{-1}
(\beta\sqrt{q}/\omega\sqrt{q-1})}\,.
\ee
Notice that while $E$ and $Q$ are proportional to $g^{-2}$, their ratio $E/Q$ 
is independent of $g$. 
A plot of $E/Q$ as a function of  $\omega,q$ for $\kappa=1$  is shown in Fig.~\ref{EQk1}.  
\subsection{Solitary Wave Bound States in the  ($\kappa$-$q$) plane}

\begin{figure}  
    \centering
    \includegraphics[width=0.7 \linewidth]{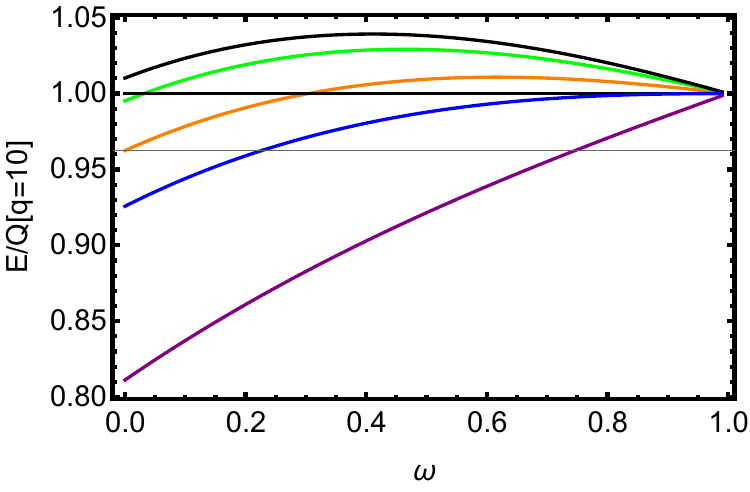}
    \caption{ \label{q10} Plot of $E/Q$ for  $q=10$  when  $\kappa=1$  (purple), $\kappa=2$ (blue), $\kappa= 2.2$ (orange), $\kappa= 2.4$ (green), $\kappa=2.5$ (black).}
\end{figure}

\begin{figure}  
    \centering
    \includegraphics[width=0.7 \linewidth]{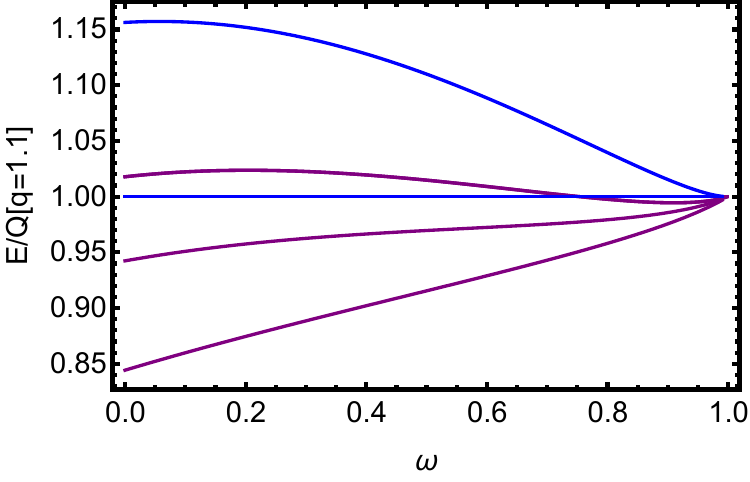}
    \caption{ \label{q1.1} Plot of $E/Q$ for  $q=1.1$  when  $\kappa=1$  (purple), $\kappa=1.2$ (red), $\kappa= 1.4$ (purple), $\kappa= 2$ (blue).}
\end{figure}

From now onwards for simplicity we use $m = 1$ so that the maximum value of
$\omega$ is one. For $\kappa \ne 1$, one has to compute $H_2$ 
and $Q$ numerically which in view of Eq. (\ref{3.18}) yields $E$ and $Q$. 
In order to have solitary wave bound states one must have $E/Q < 1$. At fixed
$q = 10$ when we go above $\kappa = 2$, we find that $E/Q$ has a maximum as a 
function of $\omega$ and at the maximum  $E/Q > 1$.  For $ 2 < \kappa < 2.5$, 
there is a region  below the maximum which shrinks with increasing $\kappa$ where  
solitary waves exist.  Once $\kappa \geq 2.5$ this region vanishes. This is displayed 
in  Fig.~\ref{q10}. From this  and other simulations at different values of $q>1$ we conclude 
that unlike the SS or the gABS cases, but like the VV case, in the complementary 
gABS model there are no solitary wave bound states in case $\kappa > 5/2$ no matter 
what $q > 1$ is. When $q=1.1$ the situation is even stricter and there are no bound 
states once $\kappa \ge 2$ as shown in Fig.~\ref{q1.1}.

\begin{figure}  
    \centering
    \includegraphics[width=0.7 \linewidth]{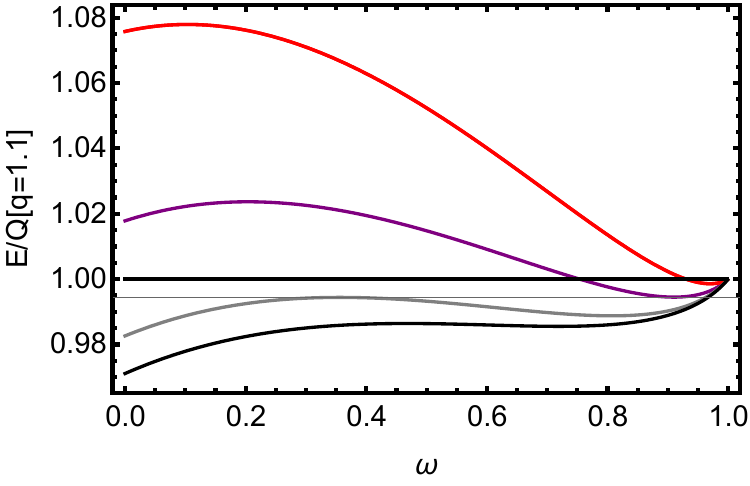}
    \caption{ \label{eovqmin} Plot of $E/Q$ for  $q=1.1$  when  $\kappa=1.27$  (black), $\kappa=1.3$ (gray), $\kappa= 1.4$ (black), $\kappa=1.6$ (red).}
\end{figure}

When $q=1.1$ there is a region between $\kappa=1.27$ and $\kappa=2$  where there is a minimum which moves toward $\omega=1$.  Already when $\kappa=1.4$ bound states exist only when $\omega > 0.75$. This region shrinks as $\kappa \rightarrow 2$.  This is seen in Fig.~\ref{eovqmin}

%
%
%

\section{Vakhitov-Kolokolov stability}

\begin{figure}  
    \centering
    \includegraphics[width=0.7 \linewidth]{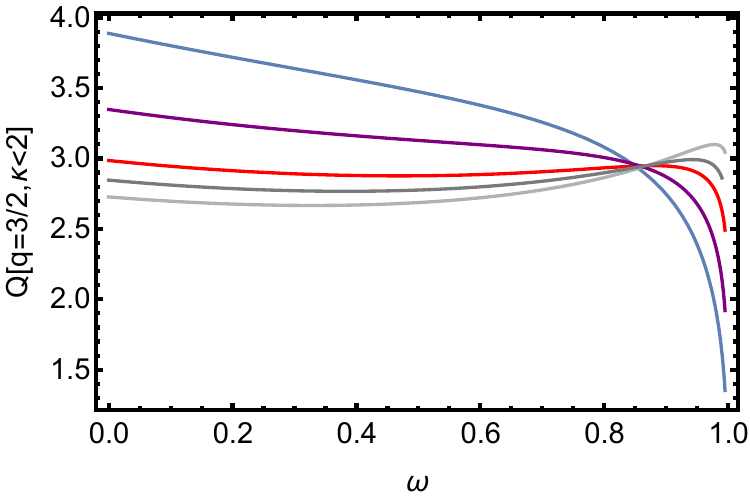}
    \caption{ \label{q32k<2} Plot of $Q(\omega) $ for  $q=3/2$  when  $\kappa=1.3$  (violet), $\kappa=1.5$ (black), $\kappa= 1.7$ (red), $\kappa= 1.8$ (gray), $\kappa=1.9$ (green).}
\end{figure}

The Vakhitov-Kolokolov stability criterion \cite{vk} is a condition for linear
stability (sometimes called spectral stability) of solitary wave solutions to a
wide class of $U(1)$-invariant Hamiltonian systems. The condition for the 
linear stability of a solitary wave 
\be\label{4.1}
u(x,t)=\phi_{\omega}(x) e^{-i\omega t}\,,
\ee
with frequency $\omega$, is given by
\be\label{4.2}
\frac{dQ(\omega)}{d\omega} < 0\,,
 \ee
where $ Q(\omega)$ is the corresponding charge which is conserved by Noether's 
theorem due to the $U(1)$-invariance of the system.
 For the VV case it is known that for $\kappa < 2$, the stability condition 
 (\ref{4.2}) is always satisfied. Remarkably, even for $\kappa >2$  it turns 
 out that there is a region $0 < \omega < \omega^{*}$ where the stability 
 condition (\ref{4.2}) is satisfied. Since in the limit $q \rightarrow \infty$,
 the above complementary generalized ABS model goes over to the VV model, 
 one would expect that the stability condition (\ref{4.2}) would be satisfied in the generalized
 model at least for the large $q$ values.  We start from $Q$ as given by Eq. (\ref{3.1})  
 where $C(\omega,\kappa)$ and
 $I(\omega,\kappa,q)$ are given by Eqs. (\ref{3.1a}) and (\ref{3.2}), respectively. Hence
 \be\label{4.3}
\frac{\partial Q(\omega,\kappa,q)}{\partial \omega} 
= \frac{\partial C(\omega,\kappa,q)}{\partial \omega} I(\omega,\kappa,q)
+ C(\omega,\kappa,q) \frac{\partial I(\omega,\kappa,q)}{\partial \omega}\,.
\ee
We find that
\be\label{4.4}
\frac{\partial C}{\partial \omega}  
= \frac{2[m-(\kappa-1)\omega][q(\kappa+1)\alpha]^{1/\kappa}}{\kappa^2 
(g^2)^{1/\kappa} \beta^{(3\kappa-1)/\kappa}}\,, 
\ee
while
\be\label{4.5}
\frac{\partial I}{\partial \omega} = \frac{dI}{d\alpha^2} \frac{d\alpha^2}
{d\omega} = - \frac{2m}{(m+\omega)^2} \frac{dI}{d\alpha^2}\,,
\ee
where
\bea\label{4.6}
&&\frac{dI}{d\alpha^2} = \frac{1}{\kappa} \int_{0}^{1} dy~ \frac{y^2}
{(1-y^2)^{(\kappa-1)/\kappa}}  \nonumber \\
&&\times  \frac{[\kappa+(\kappa+2)\alpha^2 y^2](1-\alpha^2 y^2)^{\kappa} 
- q(1+\alpha^2 y^2)^{(\kappa+1)}}{[q(1+ \alpha^2 y^2)^{\kappa+1} 
- (1- \alpha^2 y^2)^{\kappa+1}]^{(\kappa+1)/\kappa}}\,.
\eea
One actually does not have to calculate $\frac{dQ}{d \omega}$, since the sign of the slope of $Q (\omega)$ is obvious from the graphs of $Q(\omega)$.
We look at this behavior for $q=3/2$ which is far from the VV limit and at 
$q=10$ which is approaching the VV limit.   We see in the 
$q=3/2$ case the behavior of the slope changes dramatically from when $\kappa < 2$  to $\kappa >2$.  While for $\kappa < 3/2$   $Q(\omega)$ is monotonically decreasing. When   $3/2 < \kappa <2 $,  $Q$ has a minimum at  $\omega_{min}$.   We also find that   $\omega_{min}$ decreases as one increases $\kappa$. $Q$ also has a maximum in that range of $\kappa$, denoted  $\omega_{max}$ which increases in value as we increase  $\kappa <2$.  This is seen in Fig. \ref{q32k<2}.
Instead when $\kappa \ge 2$, $Q$ exhibits a shallow minimum which disappears by the time $\kappa=3$.  
\begin{figure}  
    \centering
    \includegraphics[width=0.7 \linewidth]{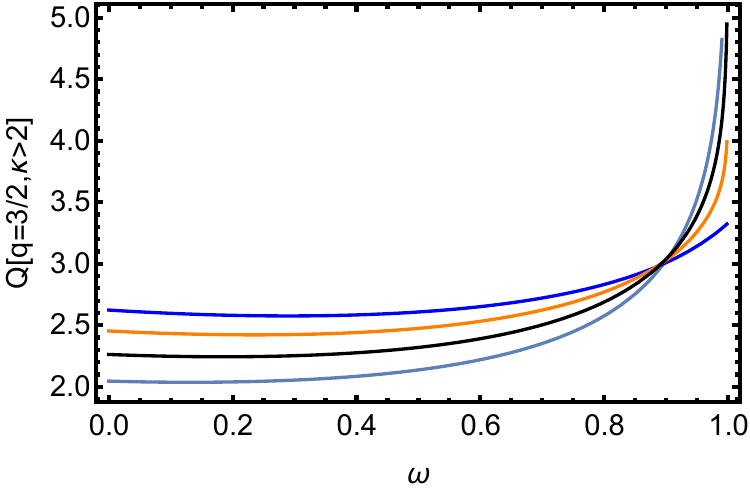}
    \caption{ \label{q32k<} Plot of $Q(\omega) $ for  $q=3/2$  when  $ \kappa=2$ (blue), $\kappa= 2.2$ (orange), $\kappa= 2.5$ (black), $\kappa=3$ (gray).}
\end{figure}
When $q=10$ one is approaching the VV case.  At this value of $q$, there  is a crossover once one passes $\kappa=2$.   For $\kappa \leq 2$,
$Q(\omega)$ is a monotonically decreasing function of $\omega$.  However once $\kappa >2$, $Q(\omega)$ develops a minimum at $\omega^*(\kappa)$ and
$\omega^*$ decreases as one increases $\kappa$.  This is seen in Fig.  \ref{Qq10}. 
\begin{figure}  
    \centering
    \includegraphics[width=0.7 \linewidth]{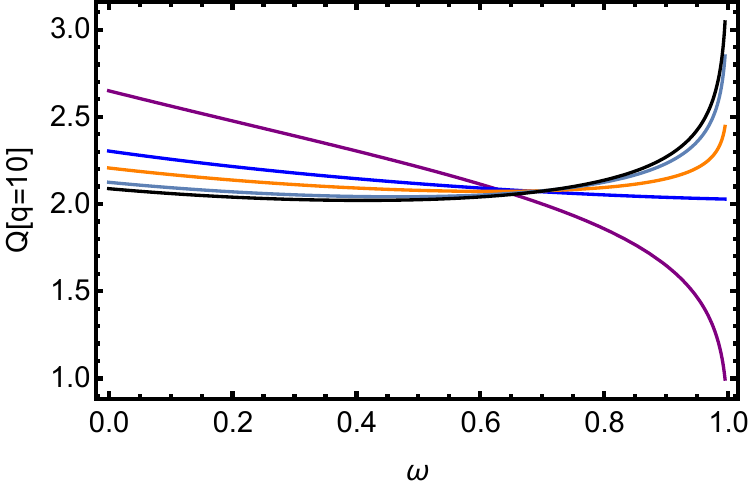}
    \caption{ \label{Qq10} Plot of $Q(\omega) $ for  $q=10$  when $ \kappa=1.5$ (purple), $\kappa=2$ (blue), $\kappa= 2.2$ (orange), $\kappa= 2.4$  (gray), $\kappa=2.5$ (black).}
\end{figure}

\section{Non-relativistic Reduction}

\begin{figure}  
    \centering
    \includegraphics[width=0.7 \linewidth] {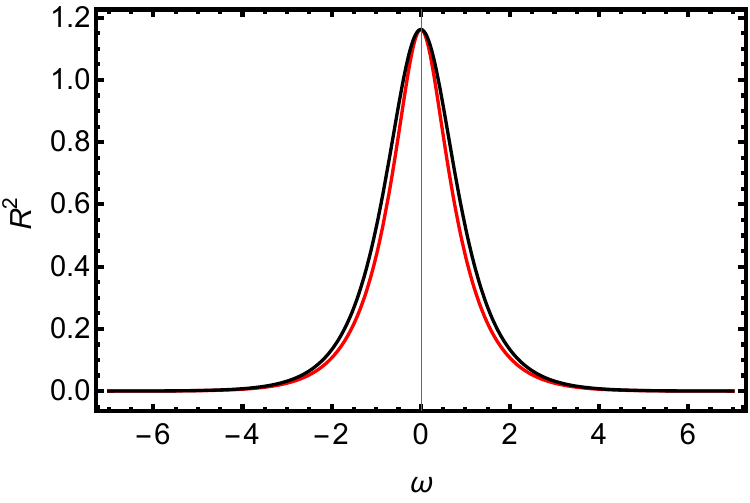}
    \caption{ \label{Rsq}  Plot of $R^2$ (red)  and $\rho$ (black)  for  $q=3$, $ \kappa=2$, and $\omega=0.7$.  }
\end{figure}

In \cite{fred} it was shown that to the leading order in $(m-\omega)/(2m)$ 
both the SS and VV interactions lead to the same modified nonlinear 
Schr\"odinger equation (NLSE).  Namely letting 
\be\label{5.1}
 \psi(x,t) = (u(x), v(x)) \, e^{-i \omega t} \>,
\ee
one finds that to the leading order, for both the SS and VV cases $u(x)$ 
obeys the NLSE given by:
\be\label{5.2}
-(u)_{xx} + (m^2 -\omega^2)u - g^2 (m +\omega)|u|^{2\kappa} u = 0 \>,
\ee
which has the exact solution,
\be\label{5.3}
u(x) = A \sech^{1/\kappa}(\kappa \beta x)\,,~~A^{2\kappa} 
= \frac{(\kappa+1)(m-\omega)}{g^2}\,.
\ee
Following similar steps, the non-relativistic version of the 
generalized present model leads to:
\be\label{5.4}
-(u)_{xx} +(m^2-\omega^2)u - \frac{(q-1)}{q} g^2 (m + \omega)|u|^{2\kappa} u 
= 0 \>,
\ee
so the only effect of the additional VV coupling is the change of the effective
coupling constant, i.e. $g^2 \rightarrow \frac{(q-1)g^2}{q}$. As a result, even
in the generalized ABS case, the same solution (\ref{5.3}) is still valid 
except $A^{2\kappa}$ is now given by
\be\label{5.5}
 A^{2\kappa} = \frac{q}{(q-1)}\frac{(\kappa+1) (m-\omega)}{g^2} \>.
\ee
Thus the non-relativistic density in the generalized ABS model is given by
\be\label{5.6}
\rho_{{NR}}(x) = \left [\frac{q}{(q-1)}\frac{(\kappa+1)(m-\omega)}{g^2}
\right]^{1/\kappa} \sech^{2/\kappa}(\kappa \beta x) \>. 
\ee
This approximation is quite good since all the relativistic solitary waves are single humped.  For example if we choose $q=3,\kappa=2,\omega=0.7$ (and also take $g=m=1$), we find the comparison  of $R^2$ in red and $\rho_{NR}$ in black shown in Fig.~\ref{Rsq}. 

\subsection{Stability in the non-relativistic regime}

As shown above, for the generalized ABS case, to the leading order in 
$(m-\omega)/(2m)$ the only effect of the parameter $p$ is to 
change the coupling constant.  However if we go to the next order then the 
Hamiltonian of the modified NLSE gets changed. For example, as shown in \cite{fred}, 
for the SS case the modified Hamiltonian is given by
\be\label{5.7}
 H_{SS} = \int \frac{dx}{2m}[\psi^{*}_x \psi_x (1 
 + \frac{\hat{g}_{s}^2}{2m}(\psi^{*} \psi)^\kappa)] 
- \frac{\hat{g}_s^2}{\kappa+1} (\psi^{*} \psi)^{\kappa+1}\,, 
\ee
whereas for the VV case the modified Hamiltonian is given by
\be\label{5.8}
 H_{VV} = \int \frac{dx}{2m}[\psi^{*}_x \psi_x (1 
 - \frac{\hat{g}_{v}^2}{2m}(\psi^{*} \psi)^\kappa)] 
- \frac{\hat{g}_{v}^2}{\kappa+1} (\psi^{*} \psi)^{\kappa+1}\,. 
\ee
Note that the next order correction has different signs for
the SS and VV cases. On following the same procedure, one finds that in the
generalized ABS case, in the next order one does not just get a redefinition of
the coupling constant.  Instead one has:
\be\label{5.9}
 H_{gABS} = \int \frac{dx}{2m}\left[\psi^\star_x \psi_x (1 
 - \frac{(1+1/q)\hat{g}_{abs}^2}{2m}(\psi^\star \psi)^\kappa)\right] 
- \frac{(1-1/q)\hat{g}_{abs}^2}{\kappa+1} (\psi^\star \psi)^{\kappa+1}\,.
\ee
Since $q > 1$, hence in the present generalized model, the non-relativistic 
Hamiltonian is the sum of two negative and one positive term: 
\be\label{5.10}
H_{gabs} = H_1 - H_2 - H_3\,. 
\ee
In the non-relativistic regime where the modified NLSE is valid we can use Derrick's 
theorem \cite{derrick} to study the stability with respect to the scale 
transformations keeping the mass 
\be\label{5.12}
M  = \int_{-\infty}^{\infty} dx\, |\psi|^2\,,
\ee
of the soliton fixed.  It is well known that for the NLSE, this method is a 
reliable tool in determining the regions of instability.  Derrick's theorem 
states that if we make the transformation
\be\label{5.13}
   \psi(x) \rightarrow \beta^{1/2} \psi(\beta x) \>,
\ee
which preserves the normalization, then if the energy is at a minimum, the 
system is stable.  The effect on the energy is then given by:
\be\label{5.14}
 H(\beta) = \beta^2 H_1 -\beta^{2+\kappa} H_2 - \beta^{\kappa} H_3 \>.
\ee
Setting the first derivative at $\beta=1$ yields:
\be\label{5.15}
   2 H_1 -(2+\kappa) H_2 - \kappa H_3 = 0 \>.
\ee
To determine when $\beta=1$ is a local minimum, we consider the second 
derivative at $\beta=1$, which is given by
\bea\label{5.16}
&&\frac{d^2H}{d\beta^2}\Big|_{\beta=1} = 2 H_1 + (\kappa+1)(\kappa+2) H_2 
- \kappa(\kappa-1) H_3  \nonumber \\
&& = 2(2-\kappa) H_1 -2(2+\kappa) H_2 \>.
\eea

Since $H_2$ is positive (and small), we conclude that $\kappa$ is slightly less 
than two, when the non-relativistic approximation is valid (i.e. when 
$(m-\omega)/(2m) \ll 1)$, the system should be stable.  

\subsection{Blowup and critical Mass}

For the NLSE the V-K condition for stability is:
\be\label{5.17}
\frac{dM(\omega)}{d\omega} > 0\,,
\ee
where $M$ is as given by Eq. (\ref{5.12}). For our generalized NLSE using 
Eq. (\ref{5.6}) we obtain
\be\label{5.18}
 M(\omega) = \frac{\sqrt{\pi}\Gamma(1/\kappa)}{\kappa \Gamma(1/2+1/\kappa)\beta}
\left[\frac{(\kappa+1)q (m -\omega)}{g^2(q-1)}\right]^{1/\kappa} \>,
\ee
so that
\be\label{5.19}
\frac{dM(\omega)}{d\omega} = -\frac{\sqrt{\pi}\Gamma(1/\kappa)[m
+(1-\kappa)\omega]}{\kappa^2 \Gamma(1/2+1/\kappa)\beta^3}
\left[\frac{(\kappa+1)q(m -\omega)}{g^2(q-1)}\right]^{1/\kappa} \>.
\ee
Thus the V-K stability will be satisfied if
\be\label{5.20}
 \omega (\kappa - 1) < m = \omega(1 + \delta)\,,
\ee
i.e. provided
\be\label{5.21}
\kappa < 2 + \delta \>.
\ee
Here $\delta = (m -\omega)/\omega$, which is assumed small. Thus as in the 
relativistic case, even in the non-relativistic case we find that the solitary 
waves are stable for $\kappa < 2$ and arbitrary $q \, (>1$).  

One can follow Section 5 of  \cite{fred} and use a self-similar 
variational approach as in that reference. One will again find that even for the 
generalized ABS case, when $\kappa=2$ and arbitrary $q \, (> 1)$, the solutions
can blowup once the initial mass of the soliton for the mNLSE is greater than a
critical value.

\section{Conclusions and Open Problems}

In this paper we have considered a two-parameter family of complementary gABS 
models characterized by the nonlinearity parameter 
$\kappa > 0$ and an admixture of SS vis a vis VV parameter $q > 1$, and have 
obtained their solitary wave solutions in the entire $(\kappa,q)$ plane. We 
showed that unlike the gABS model in this case the frequency $\omega$ is not 
restricted but can take any value $0 < \omega < m$. Further, we showed that 
unlike the gABS model \cite{ak} where the shape of the solitary wave changes
from double hump to single hump as we increase $\omega$, in the complementary 
gABS model the solitary wave always has a single hump. We also showed that 
the most contrasting behavior between the gABS and the complementary gABS 
models is in the region where solitary wave bound states exist. In particular, 
whereas in the gABS model the solitary wave bound states exist in the entire 
($\kappa$-$p$) plane ($\kappa > 0, \, p>1$), in the complementary gABS model such 
bound states exist only if $\kappa < 2.5$ no matter what $q, \omega$ are 
($q > 1, 0 < \omega < m$). We further showed that when $q = 1.1$ the solitary 
wave bound states are allowed only if $\kappa < 2$. Using Vakhitov-Kolokolov criterion  
\cite{vk} we have further shown that these solutions are stable for $\kappa < 2$. 
Finally, we have also considered the non-relativistic reduction of the complementary 
gABS model and shown that the corresponding generalized modified NLSE solutions 
are stable in case $\kappa < 2$.

This paper raises several questions, some of which are the following. 

\begin{enumerate}

\item The most important and perhaps the obvious question is about the 
stability of these solutions. One would like to know the parameter range in
the ($\kappa$-$q$) plane for which the solitary wave bound state solutions are 
stable. 

\item ABS \cite{abs} have also considered the PT-invariant generalization of 
their model. The interesting question is whether one can also construct 
the PT-invariant variant of the complementary gABS model, and if yes, can one 
also obtain its exact solutions.

\item There has been a lot of discussion in the literature about the behavior 
of the SS as well as VV solitary waves in various external fields \cite{mertens,
fred1}. It is then natural to examine the behavior of solitary waves of 
the complementary gABS model under various external fields.

\item In this paper we have considered two-parameter families of models with
VV {\it minus} $(1/q)$ SS interaction and have shown that the solitary wave bound
states exist only if $\kappa < 2.5$ irrespective of the value of $q > 1$, 
while in case $q = 1.1$ the condition gets stricter and solitary wave bound 
states exist only if $\kappa < 2$.
Another interesting model is VV {\it plus} $(1/q)$ SS interaction ($q > 0$). There 
are two interesting questions here. Firstly do bound state solitary waves now
exist in the entire ($\kappa$-$q$) plane? Secondly, for $\kappa =1$ the massive 
Thirring model (i.e. VV with $q = \infty$) is known to be integrable. Does the
integrability persist for large but finite $q$? 

\item In this paper we have considered two-parameter families of models with
a novel admixture of the VV and SS interactions. Another possible interaction
is the PS-PS (pseudoscalar-pseudoscalar) interaction. Can one construct a NLD 
model with the admixture of the PS-PS with the VV and/or with the SS models? 
Going further, can one construct a NLD model with an admixture of all three 
(i.e. SS, VV and PS-PS) interactions?

\item Finally, perhaps the most interesting question is if one can find 
relevance of the complementary gABS model (at least for one of the allowed $q$ 
values) in the context of some physical phenomena, for example in the Bose-Einstein 
condensate or photonics or condensed matter or some other physical system.
		
\end{enumerate}

We hope to address some of these questions in the near future.

\acknowledgments
We would like to thank Andrew Comech and Esthasios Charalampidis for useful 
discussions about stability.  One of us (AK) is grateful to the Indian National Science 
Academy (INSA) for the award of the INSA Honorary Scientist position at Savitribai 
Phule Pune University, India. The work at Los Alamos National Laboratory was carried 
out under the auspices of the U.S. DOE and NNSA, under Contract No. 89233218CNA000001. 

%
%

\end{document}